\documentclass[a4,11pt]{article}
\usepackage{epsfig}

\usepackage{amssymb,epsfig,graphicx,epsf}

\usepackage{amsfonts}
\usepackage{amsmath}
\usepackage{amssymb}
\usepackage{feynmf}
\newcommand{\be}{\begin{equation}}
\newcommand{\ee}{\end{equation}}

\newcommand{\ba}{\begin{eqnarray}}
\newcommand{\ea}{\end{eqnarray}}
\newcommand{\bal}{\begin{aligned}}
\newcommand{\eal}{\end{aligned}}

\begin{document}

\title{Spontaneous generation of geometry in four dimensions}

\author{Jorge Alfaro\\
Pontificia Universidad Cat\'olica de Chile, Av. Vicu\~na Mackenna 4860, Santiago, Chile,\\ 
\\
Dom\`enec Espriu and Daniel Puigdom\`enech\\
Departament d'Estructura i Constituents de la Mat\`eria,\\ 
Institut de Ci\`encies del Cosmos (ICCUB)\\ Universitat de Barcelona\\
Mart\'\i ~i Franqu\`es, 1, 08028 Barcelona, Spain.}

\date{}

\maketitle

\begin{abstract}
We present the extension to 4 dimensions of an euclidean 2-dimensional model that exhibits spontaneous 
generation of a metric. In this model gravitons emerge as Goldstone bosons of a global $SO(D) \times GL(D)$ 
symmetry broken down to $SO(D)$. The 
microscopic theory can be formulated without having to appeal to any particular space-time metric 
and only assumes the pre-existence of a manifold endowed with an affine connection. 
We emphasize that not even a flat metric needs to be assumed; in this sense 
the microscopic theory is quasi-topological. The vierbein appears as a condensate of the fundamental fermions. 
In spite of having non-standard characteristics, the microscopic theory appears to be renormalizable.
The effective long-distance theory is obtained perturbatively around a vacuum that, if the background affine 
connection is set to zero, is (euclidean) de Sitter space-time. If perturbatively small connections are 
introduced on this background, fluctuations of the metric (i.e. gravitons) appear; they are described 
by an effective theory at long distances whose more relevant operators 
correspond to the Einstein-Hilbert action with a cosmological constant. 
This effective action is derived in the large $N$ limit, $N$ being the number of fermion species in the fundamental theory.
The counterterms required by the microscopic theory are directly related to the cosmological 
constant and Newton constant and their couplings could eventually be adjusted 
to the physical values of $M_{p}$ and $\Lambda$.

\begin{flushleft}
PACS: 11.15.Ex, 04.60.Rt
\end{flushleft}
\end{abstract}
\vfill
\noindent
January 2012

\noindent
UB-ECM-FP-66/12

\noindent
ICCUB-12-002

\newpage

\section{Introduction}
It has been pointed out several times in the literature (see e.g. \cite{salam,ogi,other,also}) 
that gravitons could 
be considered as Goldstone bosons of some broken symmetry. The non-renormalizability of gravity would 
therefore be analogous to the bad ultraviolet properties of effective models of hadrons, for instance. 
This is exactly the point of view that we adopt in this paper. This work is the extension to 
four dimensions of a previous analysis in two dimensions\cite{2D}. 

In our view, concrete implementations of this idea have been lacking so far (see however \cite{ru,wett}). 
By concrete proposal we mean a field theory that does not contain the graviton field as 
an elementary degree of freedom. 
It should not even contain the tensor $\eta_{\mu\nu}$ either, as this already implies
the use of some background metric and thus the notion of geometry; gravitons being fluctuations
around this flat background. Instead, one would like to see all the metric degrees
of freedom emerging dynamically, like pions appear dynamically after chiral 
symmetry breaking in QCD. We would also like the underlying theory to be
in some sense `simpler' than gravity, in particular it should be renormalizable. 
It was shown in \cite{2D} that a theory with all these characteristics can be found in two dimensions. 

The purpose of the present work is to extend the 2D model to the far more interesting case of four dimensions. 
In spite of the complications brought up by the higher dimensionality, the nice features of the 2D case 
persist. 
The amount of divergent terms remains under control, at least on shell, and the use of the 
equations of motion ensures 
that the final effective theory is precisely that of Einstein-Hilbert plus a cosmological term, complemented 
by higher dimensional terms normalized by a computable dimensionful constant.

The topic of `emergent gravity' has recently been a popular one\cite{emergent} but
with the exceptions of the earlier works of \cite{ru} and \cite{wett} we have been unable 
to find any proposal meeting our rather restricted 
criteria of not assuming any pre-existing metric structure whatsoever. 
On the other hand the proposals in \cite{ru,wett} appear to be untractable and quantitative 
results are hard or 
impossible to get. 

Traditionally a major stumbling block in the program that will be developed below is the so-called 
Weinberg-Witten theorem \cite{ww} (see also \cite{jenkins}). The apparent pathology of theories intending
to generate dynamically gauge bosons (including gravitons in this category) 
lies in the fact that the energy-momentum
tensor has to be identically zero if massless particles with spin $\ge 1$ appear 
and one insists in the energy momentum tensor being Lorentz covariant.
However, our results, while not constituting a mathematical proof, strongly indicate 
that one can indeed get in 4D an effective low-energy theory with massless composite gravitons, so it is
legitimate to ask why Weinberg and Witten theorem would not apply. We note something peculiar
in our model, namely the energy-momentum tensor (derived in \cite{2D} for 2D and which is exactly the same
in 4D) does not have tangent (Lorentz) indices. In fact Lorentz indices are of an 
internal nature in the present approach as we will see below. 
The connection between Lorentz and world indices appears only after a vierbein is dynamically
generated. But then one is exactly in the same situation as General Relativity where the
applicability of \cite{ww} is excluded.

The dynamical generation of geometry, combined with the usual renormalization group 
arguments have rather interesting consequences.  Geometry and
distance are induced rather than fundamental concepts. At sufficiently short scales, when the
effective action does not make sense anymore, the physical degrees of freedom are 
fermionic. Below that scale there is not even the notion of distance: in a sense 
that is the shortest scale that can exist. This precludes the existence of an ultraviolet 
fixed point advocated by some \cite{fix} but also indicates that at short distances gravity 
is non-Wilsonian, suggested by others \cite{nowilson} in an holographic context. 

This paper is organized as follows: in Section 2 we review the similarities and differences 
between the 2D and 4D theories and discuss possible counterterms. In Section 3 we study the equations
of motion and revise the calculational set-up. Section 4 is devoted to the explicit 1-loop calculation 
of the one, two and three point functions of the model. 
In Section 5 we summarize the divergent content of the effective theory and make use 
of the equations of motion to prove the on-shell renormalizability of the model at leading order. 
In Section 6 the final effective action is written down and the connection between 
the free constants of the theory and the physical universal constants of gravity is discussed. 
Finally in Section 7 we summarize our results and discuss possible extensions.

\section{2-dimensional model review and 4 dimensional extension}

We showed in \cite{2D} that a consistent and renormalizable 2D model reproducing gravity 
at long distances could be built. The same model can be considered in 4D. The free Lagrangian density is
\be \label{2dlag}
\mathcal{L}_0=i\bar{\psi}_{a}\gamma^{a}\left(\partial_{\mu}+iw_{\mu}^{bc}\sigma_{bc}\right)\chi^{\mu}
+i\bar{\chi}^{\mu}\gamma^{a}\left(\partial_{\mu}+iw_{\mu}^{bc}\sigma_{bc}\right)\psi_{a},
\ee
where $\psi_{a}$ and $\chi^{\mu}$ are two species of fermions transforming, respectively, 
under Lorentz ($a,b$... are tangent space indices) and Diffeomorphisms ($\mu,\nu$... are world indices). 
A spin connection is added to the derivative to preserve the Lorentz$\times$\textsl{Diff} 
symmetry\footnote{We actually use euclidean conventions but still refer to $SO(D)$ as 
Lorentz symmetry. Note that (\ref{2dlag}) is not the usual Dirac coupling of fermions to a connection (that requires 
use of a metric). The field $\chi^\mu$ in general has a spin 1/2 and 3/2 components in general, although this statement
really makes no sense until a metric is defined.} under local 
coordinate transformations. 
It is important to notice that there is no metric to start with; there is no need of one as long 
as $\chi^{\mu}$ transforms as a spinorial density. The interaction 
term in the model, in euclidean conventions, is provided by
\be \label{interaction}
\mathcal{L}_I= iB_{\mu}^{a}(\bar{\psi}_{a}\chi^{\mu}+\bar{\chi}^{\mu}\psi_{a})+
c~\text{det}\left(B_{\mu}^{a}\right),
\ee
which obviously does not require any metric to be formulated either.
We will assume that we have $N$ species of the previous fermions but we will not add an additional 
index to avoid
complicating the notation. Note that for all practical purposes, Lorentz symmetry is an internal symmetry 
at this point.
 
The object of the interaction (\ref{interaction}) is to trigger the spontaneous breaking of 
the global symmetry via fermion condensation. Upon
use of the equations of motion for the auxiliary field $B_\mu^a$
\be \label{con}
\bar{\psi}_{a}\chi^{\mu}+\bar{\chi}^{\mu}\psi_{a}=
-ic~\frac{1}{(D-1)!}\epsilon_{aa_{2}...a_{D}}\epsilon^{\mu\mu_2...\mu_D}B_{\mu_2}^{a_2}
...B_{\mu_D}^{a_D}
\ee
and thus
\be \label{con2}
\langle \bar{\psi}_{a}\chi^{\mu}+\bar{\chi}^{\mu}\psi_{a}\rangle\neq 0\longleftrightarrow B_{\mu}^{a}\neq 0. 
\ee
If a non-zero value for the fermion condensate appears the field $B_\mu^a$ acquires an expectation value.
Such condensation was seen to happen in 2D (in the large $N$ limit) and it will be present in 
4D even for finite values of $N$ as we will see in a moment. 
Small perturbations above this vacuum expectation value will yield the effective theory of 
quantum excitations of the theory.
Our approach will be a perturbative one and it will correspond to a weak field expansion around 
the solution for $w_{\mu}^{ab}=0$. 

On translational invariance grounds, for $w_{\mu}^{ab}=0$ the vacuum of the theory is obtained 
from the gap equation for the potential
\be \label{pot}
V_{eff}=
c~\text{det}(B_{\mu}^{a})-2N\int\frac{d^Dk}{(2\pi)^D}\text{tr}(\text{log}(\gamma^{a}k_{\mu}+iB_{\mu}^{a})).
\ee
Note that the $2N$ preceding the integral comes from the $2N$ species of fermions present. 
Deriving (\ref{pot}) w.r.t. $B_{\mu}^{a}$ we obtain
\be \label{gapeq}
c~\frac{D}{D!}\epsilon_{aa_{2}...a_{D}}\epsilon^{\mu\mu_2...\mu_D}B_{\mu_2}^{a_2}
...B_{\mu_D}^{a_D}-2Ni~\text{tr}\int\frac{d^Dk}{(2\pi)^D}(\gamma^{a}k_{\mu}+iB_{\mu}^{a})^{-1}|_{a}^{\mu}=0.
\ee
This equation has a general non-trivial solution corresponding to $B_{\mu}^{a}=M\delta_{\mu}^{a}$ 
(or any $SO(D) \times GL(D)$
global transformations of this). This is analogous to the more familiar phenomenon of chiral 
symmetry breaking in strong interactions and any value of $B_\mu^a$ in the $SO(D)\times GL(D)$ 
orbit is equivalent. For simplicity we will take $B_{\mu}^{a}=M\delta_{\mu}^{a}$
and then the gap equation reduces to an equation for $M$. In 4D
\be \label{gapeq2}
\bal
cM^3-2N\int\frac{d^Dk}{(2\pi)^D}\frac{M}{k^2+M^2}=&0\\
cM^3+N\frac{M^3}{8\pi^2}\left(\frac{2}{\epsilon}-
\log{\frac{M^2 }{4\pi\mu^2}}-\gamma +1\right)=&0,
\eal
\ee
whose formal solution is
\be \label{sol}
M^2=\mu^2e^{8\pi^2 c(\mu)/N},
\ee
where $\mu\frac{dc}{d\mu}=-\frac{N}{4\pi^ 2}$, making $M$ a renormalization-group invariant. 
In the previous, we introduced the usual mass scale $\mu$ to preserve the correct 
dimensionality of the $D$-dimensional integral as dimensional regularization is used. 
For the solution to actually exist we have to require $c>0$ if $M>\mu$. If $\mu > M$ the solution exists
only if $c<0$. Therefore $c>0$ will be the case we are interested in on physical grounds. 

Note that $B_{\mu}^{a}$ has the right structure to be identified as the vierbein, 
and as it was shown in \cite{2D}, it 
consistently reemerges in the 2D effective theory to form the determinant of the spontaneously 
generated metric. 

The free fermion propagator of the theory in the broken phase can then be easily found after replacing
$B_{\mu}^{a}$ by its vacuum expectation value. With a 4D matrix notation
\be \label{prop}
\Delta^{-1}(k)_{~j}^{i}=\frac{-i}{M}\left(\delta_{~j}^{i}-\frac{\gamma^{i}(\not\! k - iM )k_j}
{k^2 + M^2}\right).
\ee

A particularity of 2D was that the most general form for $B_{\mu}^{a}$ (in euclidean conventions) 
is a conformal factor times a scale $M$ times a $\delta_{\mu}^{a}$. This means that
perturbations around the minimum of the potential can only have one physical degree of freedom, 
the conformal parameter. The other degrees of freedom in $B_{\mu}^{a}$ can be removed by suitable 
coordinate transformations and are thus unphysical (recall that the microscopic theory is fully generally 
covariant --even without a metric).

The main difference of the 4D case with respect to the 2D case is that the maximum number 
of possible physical degrees of freedom for a perturbation around the value 
$B_{\mu}^{a}=M\delta_{\mu}^a$ grows up to six instead of one, making the calculation 
much more complex. Clearly, considering a uni-parametric family of perturbations is far too simple in 4D and 
does not yield enough information to find the long-distance effective action unambiguously. To by-pass this difficulty, but 
still keeping the calculation manageable, we have chosen to restrict our considerations to diagonal
perturbations, where
\be \label{multi}
B_j^i(x) = M \delta_j^i  e^{-\frac{\sigma_i(x)}{2}} \quad({\text{ no~sum~over}}~{\it i}).
\ee  
This form contains four degrees of freedom (rather than six) but is rich enough for our purposes. 
The validity of our conclusions rely on the assumption that the effective action should be covariant 
(exactly as the microscopic theory is). This was actually
checked in the 2D case using heat kernel techniques. Here we have performed partial checks 
but we have to assume that covariance holds to draw our conclusions.

Another difference with respect to the 2D case is that the integrals involved in the perturbative calculation 
have potentially a much worse ultraviolet behavior in 4D. We postpone to Sections 3 and 4 the explicit 
calculations that indicate that the nice characteristics found in the 2D model, in particular renormalizability, 
persist in the 4D case. However the ultimate reason for renormalizability lies in the very limited number of
counterterms that can be written without a metric (and the usual assumption that the ultraviolet behavior is unaltered
by the phenomenon of spontaneous symmetry breaking).

\subsection{Possible counterterms}

Before tackling the perturbative derivation of the effective action 
it is important to list the possible invariants one can write 
in this theory without making use of a metric. In 2D we had, as already explained, two invariants that could be
constructed without having to appeal to a metric, namely ${\mathcal L}_0$ and ${\mathcal L}_I$. The latter, upon
use of the parametrization $B_\mu^a = M e^{-\frac{\sigma}{2}}$, reduces to 
\be \label{2dg}
\frac{1}{2!}\int B_{\mu}^{a}B_{\nu}^{b}\epsilon^{\mu\nu}\epsilon_{ab}\,d^2x= \ M^2\ \int \sqrt{g}\,d^2x,
\ee
i.e. is the cosmological term. In addition, there is the curvature term which in 2D, 
which in terms of the connection is simply
$\int d^2 x~dw $, purely topological; therefore we do not expect it to appear in the perturbative calculation. 
Then, apart from the free kinetic term for the fermions, there is only 
one invariant term that can be written down without a metric. Or what it is tantamount, only one 
possible counterterm remains to absorb any divergence appearing in the perturbative calculation 
after integrating out the fermions. This fact enforces the renormalizability 
of the 2D model in the large $N$ limit in spite of the bad ultraviolet behavior of the integrals. 
This argument was supported by the explicit calculations presented in \cite{2D}.

Now in 4D we can write, in addition to ${\mathcal L}_0$ and ${\mathcal L}_I$, three more counterterms 
\be \label{4dg}
{\mathcal S}_D= 
\frac{1}{4!}\int B_{\mu}^{a}B_{\nu}^{b}B_{\rho}^{c}B_{\sigma}^{d}\epsilon_{abcd}\epsilon^{\mu\nu\rho\sigma}\, d^4 x
\ee
\be \label{4dgR}
{\mathcal S}_R= \frac{1}{2}\int R_{[\mu\nu] ab}B_{\rho}^{a}B_{\sigma}^{b}\epsilon^{\mu\nu\rho\sigma}\, d^4x 
\ee
where $R_{[\mu\nu] ab}=[\nabla_{\mu ac},\nabla_{\nu cb} ]$. 
Note that $\mathcal{S}_{D}$ corresponds to the integral over the manifold of the determinant of $B_{\mu}^a$. 
After integrating the fermion fields only ${\mathcal S}_D$ and ${\mathcal S}_R$ 
can appear if general covariance is preserved. We will denote by ${\mathcal L}_D$ and ${\mathcal L}_R$ the
respective Lagrangian densities.

Finally, there is yet another counterterm one could write without making use of a metric, namely  
the Gauss-Bonnet topological invariant in 4D, which is of ${\cal O}(p^4)$ in the usual momentum counting. 
 
We did not include the term ${\mathcal S}_R$ in our action to start with because it does not contain 
the fermionic fields. It modifies neither the equation of motion (\ref{con}) for the
auxiliary field $B_\mu^a$ nor the gap equation (\ref{gapeq2}) if
the connection $w_\mu$ is set to zero as we initially do (recall that 
we use a weak field expansion and $w_{\mu}=0$ is used to determine the vacuum). However, we see 
that ${\mathcal S}_R$ is an allowed 
counterterm in 4D and therefore
it needs to be included in the initial action.
In fact, any divergence in the theory must be reabsorbable in the two terms ${\mathcal S}_D$ and ${\mathcal S}_R$, 
as they are the only local counterterms one can write before the symmetry breaking, 
i.e. before the generation of the metric. 
 
When the auxiliary field $B_\mu^a$ is identified with the vierbein, the parametrization (\ref{multi}) 
and the equations of motion are used the
two counterterms reduce to
\be \label{conterterms}
 M^4\ \int \sqrt{g}~d^4x,\qquad M^2 \int \sqrt{g}R ~d^{4}x,
\ee
respectively; i.e. the familiar cosmological and Einstein terms. This will be explained in more detail in 
the next section.

\section{Equations of motion}

Let us write explicitly what the 4D counterterms 
look like once we replace $B_\mu^a$ by its vacuum expectation value plus perturbations around it. To keep the
notation simple, let us consider the case in (\ref{multi}) when $\sigma_i (x)= \sigma(x)$ (conformally flat metric).
For $B_{\mu}^{a}=Me_{\mu}^{a}=Me^{-\sigma/2}\delta_{\mu}^{a}$ we have
\be \label{sqrg}
\mathcal{L}_{D}= 
\frac{1}{4!}B_{\mu}^{a}B_{\nu}^{b}B_{\rho}^{c}B_{\sigma}^{d}\epsilon_{abcd}\epsilon^{\mu\nu\rho\sigma}= M^4e^{-2\sigma}
\ee
and
\be \label{R}
 \frac{1}{2}R_{[\mu\nu] ab}B_{\rho}^{a}B_{\sigma}^{b}\epsilon^{\mu\nu\rho\sigma}=\frac{1}{2}(\partial_{\mu} w_{\nu}^{~\mu\nu}-\partial_{\nu} w_{\mu}^{~\mu\nu}+ w_{\mu}^{~\mu c} w_{\nu}^{~c\nu}- w_{\nu}^{~\mu c} w_{\mu}^{~c\nu})e^{-\sigma}M^2.
\ee
Note that because $e_{\mu}^{a}=\delta_{\mu}^{a}$ we can use indistinctively greek and latin indices; 
they are lowered and raised with a trivially flat metric. This is somewhat similar to what happens in 
linearized gravity where indices are lowered and raised with $\eta_{\mu\nu}$.

Let us now work out the equations of motion for the full Lagrangian $\mathcal{L}_{0}+\mathcal{L}_{I}+\mathcal{L}_{R}$. In Section 2 we already discussed the equations of motion for the field $B_{\mu}^{a}$ when $w_{\mu}^{ab}=0$. In addition we have 
\be \label{eom}
\begin{aligned}
\frac{\delta \mathcal{L}}{\delta w_{\mu}^{~ab}}
=&\partial_{\rho}\left(\frac{\delta (\mathcal{L}_{0}+\mathcal{L}_{I}+\mathcal{L}_{R})}{\delta \partial_{\rho} w_{\mu}^{~ab}}\right)-\frac{\delta (\mathcal{L}_{0}+\mathcal{L}_{I}+\mathcal{L}_{R})}{\delta w_{\mu}^{~ab}}=0\\
=&\frac{1}{2}\left(-\partial_{a}\sigma \delta_{b}^{\mu}+\partial_{b}\sigma \delta_{a}^{\mu}-\delta_{a}^{\mu} w_{\nu b}^{~~\nu}-\delta_{b}^{\mu} w_{\nu~ a}^{~\nu}+ w_{ab}^{~~\mu}+ w_{b~a}^{~\mu}\right)e^{-\sigma}\\
&-\frac{1}{M^2}(\bar\psi_{c}\gamma^{c}\sigma_{ab}\chi^{\mu}+\bar{\chi}^{\mu}\gamma^{c}\sigma_{ab}\psi_{c})=0.
\end{aligned}
\ee
To solve (\ref{eom}) we will only consider the lowest order term in the $1/M^2$ expansion, following the usual counting
rules in effective Lagrangians based on a momentum expansion.
The general solution for the connection is then
\be \label{eom2}
 w_{\mu}^{~ab}=\frac{1}{2}(\partial^{a}\sigma\delta_{\mu}^{b}-\partial^{b}\sigma\delta_{\mu}^{a}).
\ee
This is precisely the usual condition between the spin connection 
and the vierbein in General Relativity (\ref{spin}), characteristic of the Palatini formalism\cite{pal}
\be \label{spin}
w_{\mu}^{ab}=e_{\nu}^{a}\partial_{\mu}E^{\nu b}+e_{\nu}^{a}E^{\rho b}\Gamma_{\mu\rho}^{\nu}
\ee
particularized to a conformally flat metric given by $e_{\mu}^{a}=\delta_{\mu}^{a}$ ($E^{\rho b}$ is the inverse
vierbein).
Making use of (\ref{eom2}) in (\ref{R}) we are now allowed to identify the curvature in terms of the scalar field $\sigma$
\be \label{Rsig}
\mathcal{L}_{R}|_{\text{(on shell)}}=
M^2\sqrt{g}R=\frac{3}{2}\left(\Box \sigma-\frac{1}{2}\partial_{\mu}\sigma\partial^{\mu}\sigma\right)e^{-\sigma}M^2
\ee
Note that in the particular case of a vierbein corresponding to a conformally flat metric 
one can integrate by parts either 
of the terms in (\ref{Rsig}) to obtain the other
\be \label{Rfin}
\begin{aligned}
\sqrt{g}R&=\frac{3}{2}M^2(\Box \sigma-\frac{1}{2}\partial_{\mu}\sigma\partial^{\mu}\sigma)e^{-\sigma}
&=\frac{3}{4}M^2(\Box \sigma)
(1-\sigma+\frac{\sigma^2}{2}-\frac{\sigma^3}{6}+...)
\end{aligned}
\ee
This term plus a constant times (\ref{sqrg}) are the only divergences that should appear in the final effective theory upon
integration of the fermionic fields for this particular type of perturbations above the vacuum (i.e. those interpretable as a
conformally flat metric).

As previously mentioned we shall consider a more general type of perturbations; namely, we will 
use the diagonal parametrization of the perturbations around the vacuum solution given by (\ref{multi}). This is not
the most general one in 4D, but it is enough for our purposes. After the identification of $B_\mu^a$ with the vierbein, this corresponds to a metric 
\be \label{g}
g_{\mu\nu}=\left(\begin{matrix}
e^{-\sigma_{1}(x)} & 0 & 0 & 0\\
0 & e^{-\sigma_{2}(x)} & 0 & 0\\
0 & 0 & e^{-\sigma_{3}(x)} & 0\\
0 & 0 & 0 & e^{-\sigma_{4}(x)}\\
\end{matrix}\right).
\ee
This parametrization provides enough generality to the calculation. 
We can now derive the equivalent expression to (\ref{eom2}) for the general diagonal 
perturbation using (\ref{spin}) to obtain \footnote{Again we emphasize that although it may seem 
strange to see latin indices in the derivatives this should not confuse the reader. 
After the symmetry breaking a vierbein is generated relating world 
indices with tangent space ones through $\delta_{\mu}^{a}$. In the expression (\ref{eom4}) we have compiled
the entries for $w_\mu^{ab}$ in a bi-matrix form, but they {\em should not} be multiplied; only the index
$\rho$ is summed up.}
\be \label{eom4}
\bal
w_{\mu}^{~a b}=&\frac{1}{2}
\left[ \left(\begin{matrix}
e^{\frac{\sigma_{1}}{2}} & 0 & 0 & 0 \\
0 & e^{\frac{\sigma_{2}}{2}} & 0 & 0 \\
0 & 0 & e^{\frac{\sigma_{3}}{2}} & 0 \\
0 & 0 & 0 & e^{\frac{\sigma_{4}}{2}} \\
\end{matrix}\right)^{a\rho}\left(\begin{matrix}
\partial_{\rho}\sigma_{1}e^{-\frac{\sigma_{1}}{2}} & 0 & 0 & 0 \\
0 & \partial_{\rho}\sigma_{2}e^{-\frac{\sigma_{2}}{2}} & 0 & 0 \\
0 & 0 & \partial_{\rho}\sigma_{3}e^{-\frac{\sigma_{3}}{2}} & 0 \\
0 & 0 & 0 & \partial_{\rho}\sigma_{4}e^{-\frac{\sigma_{4}}{2}} \\
\end{matrix}\right)^{b}_{\mu}\right. \\
& \left. -\left(\begin{matrix}
e^{\frac{\sigma_{1}}{2}} & 0 & 0 & 0 \\
0 & e^{\frac{\sigma_{2}}{2}} & 0 & 0 \\
0 & 0 & e^{\frac{\sigma_{3}}{2}} & 0 \\
0 & 0 & 0 & e^{\frac{\sigma_{4}}{2}} \\
\end{matrix}\right)^{b\rho}\left(\begin{matrix}
\partial_{\rho}\sigma_{1}e^{-\frac{\sigma_{1}}{2}} & 0 & 0 & 0 \\
0 & \partial_{\rho}\sigma_{2}e^{-\frac{\sigma_{2}}{2}} & 0 & 0 \\
0 & 0 & \partial_{\rho}\sigma_{3}e^{-\frac{\sigma_{3}}{2}} & 0 \\
0 & 0 & 0 & \partial_{\rho}\sigma_{4}e^{-\frac{\sigma_{4}}{2}} \\
\end{matrix}\right)^{a}_{\mu}\right].
\eal
\ee
Making use of the equations of motion one can compute the corresponding 
$\mathcal{L}_{R}$ for the general case and expand it 
in the $\sigma$ fields. The result up to two sigma fields reads
\be \label{Rmulti}
\bal
\mathcal{L}_{R}|_{\text{(on shell)}}=&M^2\sqrt{g}R=M^2
\left[
 \partial_{3}^{2}\sigma_{4}+\partial_{2}^{2}\sigma_{4}+\partial_{1}^{2}\sigma_{4}
 +\partial_{4}^{2}\sigma_{3}+\partial_{2}^{2}\sigma_{3}+\partial_{1}^{2}\sigma_{3}
 \right.\\
&\left.+\partial_{4}^{2}\sigma_{2}+\partial_{3}^{2}\sigma_{2}+\partial_{1}^{2}\sigma_{2}+\partial_{4}^{2}\sigma_{1}+\partial_{3}^{2}\sigma_{1}+\partial_{2}^{2}\sigma_{1}
\right.\\
&\left.
 -\frac{1}{2}\left(\partial_{3}\sigma_{1}\partial_{3}\sigma_{2}+\partial_{4}\sigma_{1}\partial_{4}\sigma_{2}
+\partial_{2}\sigma_{1}\partial_{2}\sigma_{3}+\partial_{4}\sigma_{1}\partial_{4}\sigma_{3}+\partial_{2}\sigma_{1}\partial_{2}\sigma_{4}
+\partial_{3}\sigma_{1}\partial_{3}\sigma_{4}\right.\right.\\
&\left.\left. +\partial_{1}\sigma_{2}\partial_{1}\sigma_{3}+\partial_{4}\sigma_{2}\partial_{4}\sigma_{3}+\partial_{1}\sigma_{2}\partial_{1}\sigma_{4}+\partial_{3}\sigma_{2}\partial_{3}\sigma_{4}
+\partial_{1}\sigma_{3}\partial_{1}\sigma_{4}+\partial_{2}\sigma_{3}\partial_{2}\sigma_{4}\right)\right.\\
&\left. +\mathcal{O}(\sigma^3)\right]
.
\eal
\ee
More details on the calculation of (\ref{Rmulti}) can be found in the Appendix. 
Ignoring for a moment the Gauss-Bonnet invariant, the divergent terms from the perturbative calculation 
for the general perturbation should match on shell either with (\ref{Rmulti}) or with
\be \label{gmulti}
\mathcal{L}_{D}|_{\text{(on shell)}}=M^4\sqrt{g}=M^4e^{-\frac{\sum_i \sigma_{i}}{2}}.
\ee

An extension to the most general perturbation with the full six degrees of freedom should 
be possible but would require much more effort,
which we consider unnecessary at this point as the above parametrization provides enough redundancy. 
Since the coefficients for the terms in the effective action are universal there should be 
no loss of generality in the present approach.
This of course assumes that general covariance is kept all along the derivation of the effective action
and by the regulator, as it should be the case in dimensional regularization.

So far we have explained how the 2D model can be consistently extended to 4D preserving the 
key features. We study small perturbations around a constant vacuum expectation value for the field $B_{\mu}^{a}$ 
(which does not need to be small itself) corresponding to the solution of the gap equation for $w_{\mu}^{ab}=0$. 
In such a theory one can write a limited number of counterterms without making use of a metric. 
These counterterms are consistent with the usual terms of GR once used the equations of motion. 
With all these ingredients we are ready to move to the actual perturbative derivation of the effective action.

\section{1-loop structure for a general diagonal perturbation}

The effective action that describes perturbations above the trivial vacuum 
\be
w_\mu^{ab}=0,\qquad  B_\mu^a= M \delta_\mu^a,
\ee
will be given by a polynomial expansion in powers of $w_\mu (x)$, $\sigma_i(x)$ and their derivatives
obtained after integration of the fundamental degrees of freedom. In this 
section we will derive this effective action diagrammatically.

We shall use the diagonal perturbation (\ref{multi}) with 4 degrees of freedom for the vierbein perturbations. 
For simplicity, we will calculate 
only the one-point and two-point functions for this rather general case and then particularize to the conformal 
case ($\sigma_i(x)=\sigma(x)$) to compute some three-point functions. 

Since perturbation theory in this model has some peculiar features (note in particular the behavior of the fermion propagator) in what follows we shall provide enough details so that the diagrammatic calculation can be reproduced.

Starting from the Lagrangian density ${\mathcal L}_0 + {\mathcal L}_I$ described in Section 2 (note that
${\mathcal L}_R$  plays no role whatsoever in the integration of the $N$ species of fermions), and using a
parametrization of  $B_{\mu}^{a}$ given by (\ref{multi}), the interaction vertices are
\begin{figure}[h]
\begin{equation}\label{vert11}
\begin{aligned}
\parbox{30mm}{
\includegraphics[]{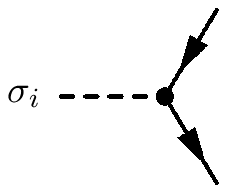}}
&\qquad i\frac{1}{2}M\delta_{~\mu}^{i}\\
\\
\parbox{30mm}{
\includegraphics[]{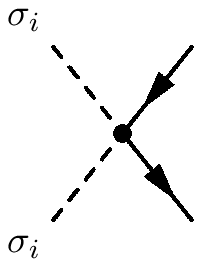}}
&\qquad -i\frac{1}{8}M\delta_{~\mu}^{i}\\
\\
\parbox{30mm}{
\includegraphics[]{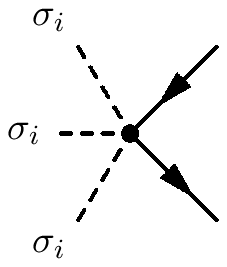}}
&\qquad i\frac{1}{48}M\delta_{~\mu}^{i}\\
\end{aligned}
\end{equation}
\end{figure}

\subsection{One and two point functions for the fields $\sigma_{i}$}
With the rules described above and using the propagator (\ref{prop}) we can calculate the first 
1-loop diagrams for $D=4-\epsilon$. We will not include the factor $N$ in the diagrammatic results
presented below. The vacuum bubble diagram is\\
\\
\be \label{S0}
\bal
&~~~~~~~~~~~~~~~~~~~~~~~~~~~~~~~~
\parbox{80mm}{
\includegraphics[]{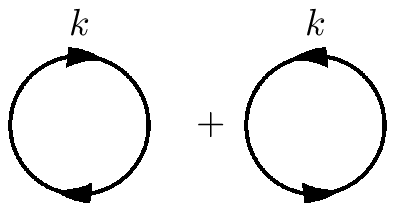}}
\\
&=-2\text{Tr}\left[\int \frac{d^Dk}{(2\pi)^D}\Delta^{-1}(k)_{~a}^{\mu}(-i)\delta_{~\mu}^{a}\right]
=-\frac{M^3}{2\pi^2}\left(\frac{2}{\epsilon}-\log{\left(\frac{M^2}{4 \pi\mu^2 }\right)}-\gamma +1\right).\\
\eal
\ee
We also compute the one-point function for the different vertices
\be \label{S01}
\bal
&~~~~~~~~~~~~~~~~~~~~~~~~~
\parbox{80mm}{
\includegraphics[]{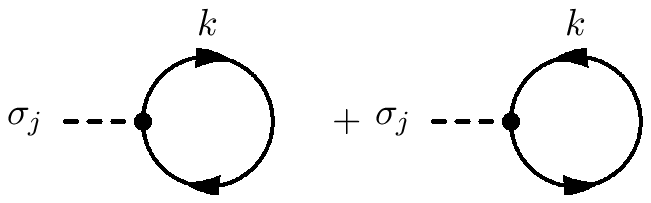}}
\\
&=-\sum_{j=1}^4\sigma_{j}\frac{2}{2}\text{Tr}\left[\int \frac{d^Dk}{(2\pi)^D}\Delta^{-1}(k)_{~j}^{\mu}(i)M\delta_{~\mu}^{j}\right]
=\sum_{j=1}^4\frac{\sigma_{j}M^4}{16\pi^2}\left(\frac{2}{\epsilon}-\log{\left(\frac{M^2}{4 \pi\mu^2 }\right)}-\gamma +1\right),\\
\\
\eal
\ee
\be \label{S02}
\bal
&~~~~~~~~~~~~~~~~~~~~~~~~~~~~~~
\parbox{80mm}{
\includegraphics[]{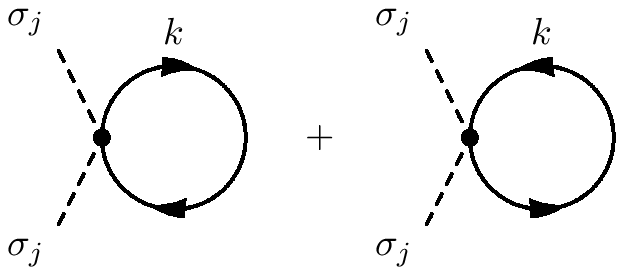}}
\\
&=-\sum_{j=1}^{4}\sigma_{j}^2\frac{2!~2}{2!~2\cdot 2}\text{Tr}\left[\int \frac{d^Dk}{(2\pi)^D}\Delta^{-1}(k)_{~j}^{\mu}(-i)M\delta_{~\mu}^{j}\right]
=-\sum_{j=1}^{4}\sigma_{j}^2\frac{M^4}{32\pi^2}\left(\frac{2}{\epsilon}-\log{\left(\frac{M^2}{4 \pi\mu^2 }\right)}-\gamma +1\right).\\
\\
\eal
\ee
Let us, for this particular diagram, clarify what the origin of the numerical factor is. In the numerator, the $2!$ 
comes from the combinatorial possible connections of the external fields $\sigma_{i}$. The other $2$ is due to the 
two species of fermions and it is present in all diagrams. In the denominator, $2! 2\cdot 2$ comes from the vertex. 
Since it is a one point function there are no additional factors, however for n-point functions the corresponding $n!$ 
will be present in the denominator.\\
\\
\be \label{S03}
\bal
&~~~~~~~~~~~~~~~~~~~~~~~~~~~~~
\parbox{80mm}{
\includegraphics[]{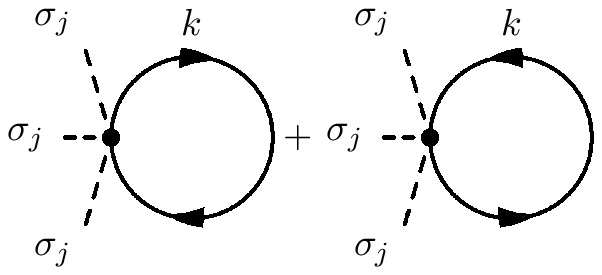}}
\\
&=-\sum_{j=1}^{4}\frac{3!~2}{3!~2\cdot 2\cdot 2}\text{Tr}\left[\int \frac{d^Dk}{(2\pi)^D}\Delta^{-1}(k)_{~j}^{\mu}(i)M\delta_{~\mu}^{j}\right]
=\sum_{j=1}^{4}\frac{\sigma_{j}^{3}M^4}{64\pi^2}\left(\frac{2}{\epsilon}-\log{\left(\frac{M^2}{4 \pi\mu^2 }\right)}-\gamma +1\right).\\
\eal
\ee
Next diagram is the two-point function
\be \label{sigsig}
\bal
&~~~~~~~~~~~~~~~~~~~~~~~~~
\parbox{80mm}{
\includegraphics[]{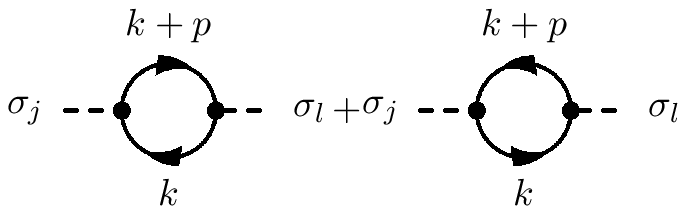}}
\\
&=-\sum_{j=1}^{4}\sum_{l=1}^{4}\sigma_{j}\sigma_{l}\frac{2!~2}{2!\cdot 2\cdot 2}\text{Tr}\left[\int \frac{d^Dk}{(2\pi)^D}iM\delta_{~\mu}^{l}\Delta^{-1}(k)_{~j}^{\mu}iM\delta_{~\nu}^{j}\Delta^{-1}(k+p)_{~l}^{\nu}\right]\\
&=\sum_{j=1}^{4}\sum_{\bal l=1\\ l\neq j\eal}^{4}\left[-\frac{\sigma_{j}\sigma_{l}M^4}{16\pi^2}\left(\frac{2}{\epsilon}-\log{\left(\frac{M^2}{4 \pi\mu^2 }\right)}-\gamma +\frac{2}{3}\right)-\frac{\sigma_{j}\sigma_{l}p^2M^2}{48\pi^2}\left(\frac{2}{\epsilon}-\log{\left(\frac{M^2}{4 \pi\mu^2 }\right)}-\gamma -\frac{1}{3}\right)\right.\\
&\left. ~~~+\frac{\sigma_{j}\sigma_{l}(p_{j}^2+p_{l}^{2})M^2}{48\pi^2}\left(\frac{2}{\epsilon}-\log{\left(\frac{M^2}{4 \pi\mu^2 }\right)}-\gamma -\frac{1}{2}\right)\right]\\
&~~~~+\sum_{j=1}^{4}\left[+\frac{\sigma_{j}^2M^4}{32\pi^2}+\frac{\sigma_{j}^2M^2p^2}{32\pi^2}-\frac{\sigma_{j}^2M^2p_{j}^2}{32\pi^2}\right]+\mathcal{O}(p^4).
\eal
\ee
The numerical factor in this case is composed by $2!~2$ in the numerator from the possible contractions of 
external fields times the two species of fermions. And $2!$ in the denominator from the diagram being a 
two-point function and finally the $1/2$ from each vertex. We will not elaborate on the combinatorial factors anymore
but we write all factors explicitly, even if the notation may be a bit cumbersome, in order to facilitate the check
of our results.
By $\mathcal{O}(p^4)$ we mean 
finite higher order in $p^2$ contributions.
\subsection{Diagrams with $w_{\mu}^{ab}$}
Now we turn to the diagrams that contain a field $w_{\mu}^{ab}$. The corresponding vertex is
\begin{equation}\label{vertW}
\parbox{30mm}{
\includegraphics[]{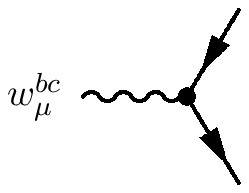}}
\\
\qquad i\gamma^{a}w_{\mu}^{bc}\sigma_{bc}=\frac{\gamma^{a}}{4}[\gamma_{b},\gamma_{c}].\\
\end{equation}

The one and two point functions yield
\be \label{W1S0}
\bal
&~~\parbox{80mm}{
\includegraphics[]{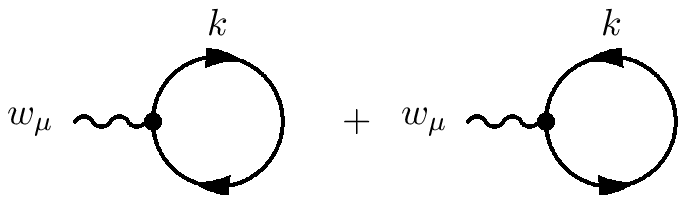}}
\\
\\
&=-2\text{Tr}\left[\int \frac{d^Dk}{(2\pi)^D}i\gamma^{a}\sigma_{bc}\Delta^{-1}(k)_{~a}^{\mu}\right]
=0\\
\eal
\ee

then
\be \label{W1S1}
\bal
&~~~~~~~~~~~~~~~~~\parbox{80mm}{
\includegraphics[]{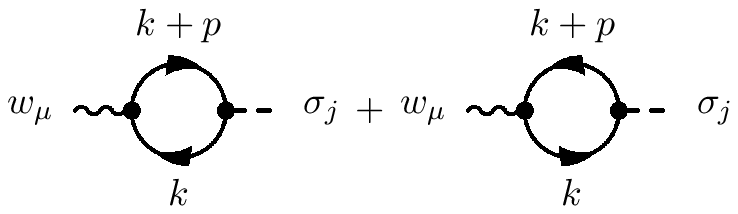}}
\\
&=-\sum_{j=1}^{4}\sigma_{j}\frac{2}{2!\cdot 2}\text{Tr}\left[\int \frac{d^Dk}{(2\pi)^D}i\gamma^{a}\sigma_{bc}\Delta^{-1}(k)_{~d}^{\mu}iM\delta_{~\nu}^{d}\Delta^{-1}(k+p)_{~a}^{\nu}\right]=0.
\eal
\ee
Suggesting that diagrams containing only one field $w_{\mu}^{ab}$ are zero. For two $w_{\mu}^{ab}$ fields we have
\be \label{W2}
\bal
&~~~~~~~~~~~~~~~~~~~~~~~~~~\parbox{80mm}{
\includegraphics[]{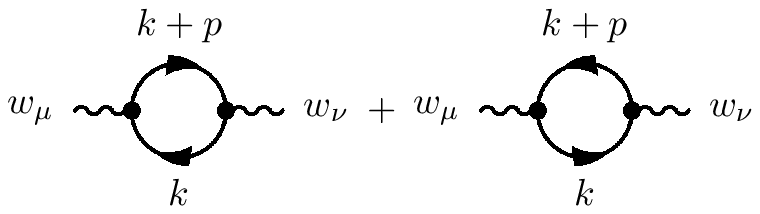}}
\\
=&-\frac{2!\cdot 2}{2!}\text{Tr}\left[\int \frac{d^Dk}{(2\pi)^D}i\gamma^{a}\sigma_{bc}\Delta^{-1}(k)_{~d}^{\mu}i\gamma^{d}\sigma_{ef}\Delta^{-1}(k+p)_{~a}^{\nu}\right]\\
=&\left(\frac{2}{\epsilon}-\log{\left(\frac{M^2}{4 \pi\mu^2 }\right)}-\gamma +\frac{1}{4}\right)\left[\frac{M^2}{4\pi^2}(\delta_{be}\delta_{c}^{\nu}\delta^{\mu}_{f}-\delta_{be}\delta_{cf}\delta^{\mu\nu}+\delta_{bf}\delta_{ce}\delta^{\mu \nu}-\delta_{bf}\delta_{c}^{\nu}\delta^{\mu}_{e}\right. \\
&\left. +\delta_{b}^{\nu}\delta_{cf}\delta^{\mu}_{ e}-\delta_{b}^{\nu}\delta_{c e}\delta^{\mu}_{ f})\right]
+\frac{M^2}{16 \pi^2 }\left(\delta_{bf}\delta_{e}^{\nu}\delta_{c}^{\mu }-\delta_{be}\delta_{c}^{\mu}\delta_{f}^{\nu}+\delta_{b}^{\mu}\delta_{ce}\delta_{f }^{\nu}-\delta_{b}^{\mu}\delta_{e}^{\nu}\delta_{cf}\right)\\
&+\frac{1}{\epsilon}F^{\mu\nu}_{~~~ bcef}(p^ 2)+\mathcal{O}(p^2)\\
=&\left(\frac{2}{\epsilon}-\log{\left(\frac{M^2}{4 \pi\mu^2 }\right)}-\gamma +\frac{1}{4}\right)\left[\frac{M^2}{4\pi^2}D^{\mu\nu}_{~~~ bcef}\right]+\frac{M^2}{16 \pi^2 }E^{\mu\nu}_{~~~ bcef}+\frac{1}{\epsilon}F^{\mu\nu}_{~~~ bcef}(p^ 2)+\mathcal{O}(p^2)\\
\eal
\ee
Where $F^{\mu\nu}_{~~~ bcef}$ is a complicated structure composed of external momenta and Kronecker 
deltas of order $\mathcal{O}(p^2)$. This divergence is of higher order, in the $1/M^2$ expansion, 
than the one of $D^{\mu\nu}_{~~~ bcef}$. Now, taking into account 
that $w_{\mu}^{bc}=-w_{\mu}^{cb}$, we can show that
\be \label{ww}
D^{\mu\nu}_{~~~ bcef}w_{\mu}^{bc}w_{\nu}^{ef}=
4w_{\mu}^{\nu b}w_{\nu}^{\mu b}-2w_{\mu}^{be}w_{\mu}^{be}=0;\quad E^{\mu\nu}_{~~~ bcef}w_{\mu}^{bc}w_{\nu}^{ef}=4w_{\mu}^{\mu b}w_{\nu}^{b \nu}
\ee
More details can be found in the appendix.
\subsection{Three-point functions}

In order to keep the calculations simple, we particularize to the case $B_{\mu}^{a}=Me^{-\frac{\sigma(x)}{2}}\delta_{\mu}^{a}$.
The previous results, (\ref{S0}-\ref{ww}) are all valid taking $\sigma_{i}=\sigma$, $i=1,2,3,4$. 
With  this simplification we can easily further compute more diagrams. For the field $\sigma$ we have
\be \label{sigsig2}
\bal
\\
&~~~\parbox{150mm}{
\includegraphics[]{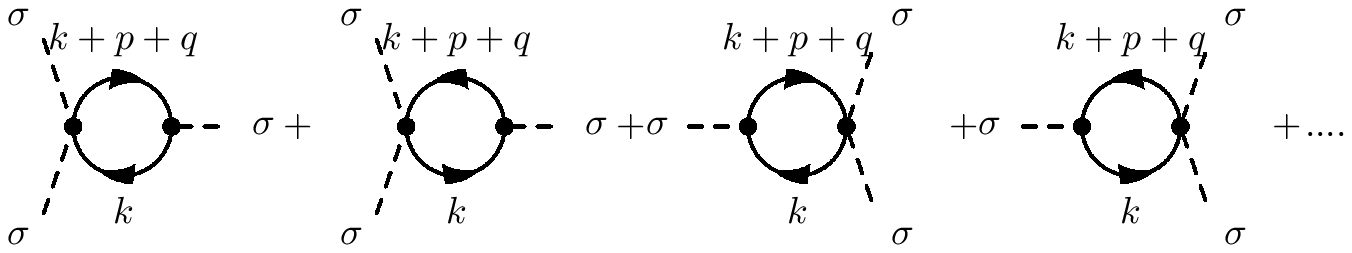}}
\\
=&-\frac{2~2~2}{2!~2!\cdot 2\cdot 2\cdot 2}\text{Tr}\left[\int \frac{d^Dk}{(2\pi)^D}(-i)M\delta_{~\mu}^{a}\Delta^{-1}(k)_{~b}^{\mu}iM\delta_{~\nu}^{b}\Delta^{-1}(k+p+q)_{~a}^{\nu}\right]\\
&-\frac{2~2~2}{2!~2!\cdot 2\cdot 2\cdot 2}\text{Tr}\left[\int \frac{d^Dk}{(2\pi)^D}(-i)M\delta_{~\mu}^{a}\Delta^{-1}(k+p)_{~b}^{\mu}iM\delta_{~\nu}^{b}\Delta^{-1}(k+p+q)_{~a}^{\nu}\right]\\
&-\frac{2~2~2}{2!~2!\cdot 2\cdot 2\cdot 2}\text{Tr}\left[\int \frac{d^Dk}{(2\pi)^D}(-i)M\delta_{~\mu}^{a}\Delta^{-1}(k+q)_{~b}^{\mu}iM\delta_{~\nu}^{b}\Delta^{-1}(k+p+q)_{~a}^{\nu}\right]\\
=&\frac{9M^4}{16\pi^2}\left(\frac{2}{\epsilon}-\log{\left(\frac{M^2}{4 \pi\mu^2 }\right)}-\gamma +\frac{1}{3}\right)\\
&+\frac{M^2(p^2+(p+q)^2+q^2)}{32\pi^2}\left(\frac{2}{\epsilon}-\log{\left(\frac{M^2}{4 \pi\mu^2 }\right)}-\gamma -\frac{2}{3}\right)+\mathcal{O}(p^4)
\eal
\ee

And also
\be \label{sig3pots}
\bal
\\
&~~~~~~~~~~~~~~~~~~~~~~~~~~~~~\parbox{80mm}{
\includegraphics[]{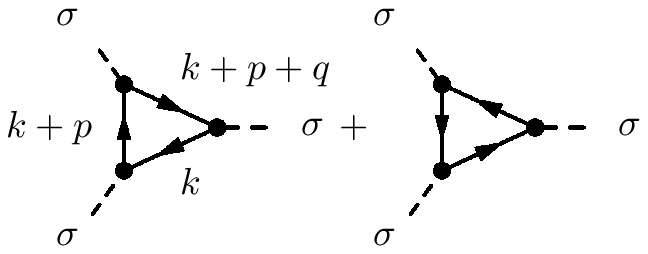}}
\\
=&-\frac{3!~2}{3!~2\cdot 2\cdot 2}\text{Tr}\left[\int \frac{d^Dk}{(2\pi)^D}iM\Delta^{-1}(k)_{~b}^{\mu}iM\delta_{~\nu}^{b}\Delta^{-1}(k+p+q)_{~d}^{\nu}iM\delta_{~\rho}^{d}\Delta^{-1}(k+p)_{~a}^{\rho}\right]\\
&-\frac{3!~2}{3!~2\cdot 2\cdot 2}\text{Tr}\left[\int \frac{d^Dk}{(2\pi)^D}iM\Delta^{-1}(k)_{~b}^{\mu}iM\delta_{~\nu}^{b}\Delta^{-1}(k+p+q)_{~d}^{\nu}iM\delta_{~\rho}^{d}\Delta^{-1}(k+q)_{~a}^{\rho}\right]\\
=&\frac{3M^4}{8 \pi^2 }\left(\frac{2}{\epsilon}-\log{\left(\frac{M^2}{4 \pi\mu^2 }\right)}-\gamma -\frac{2}{3}\right)-M^2\left(\frac{p^2}{32 \pi^2 }+\frac{(p+q)^2}{32 \pi^2 }+\frac{q^2}{32 \pi^2 }\right)+\mathcal{O}(p^4)
\eal
\ee
On the other hand, for the field $w_{\mu}^{ab}$ we can compute
\be \label{W1S2}
\bal
\\
&~~~~~~~~~~~~~~~~~~~~~~~~~~\parbox{80mm}{
\includegraphics[]{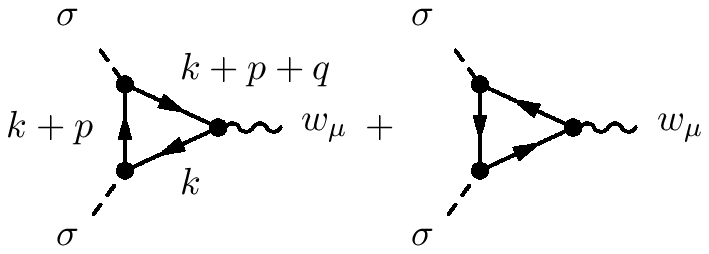}}
\\
=&-\frac{2!~2}{3!~2\cdot 2}\text{Tr}\left[\int \frac{d^Dk}{(2\pi)^D}i\gamma^{a}\sigma_{bc}\Delta^{-1}(k)_{~d}^{\mu}iM\delta_{~\nu}^{d}\Delta^{-1}(k+p+q)_{~e}^{\nu}iM\delta_{~\rho}^{e}\Delta^{-1}(k+p)_{~a}^{\rho}\right]\\
&-\frac{2!~2}{3!~2\cdot 2}\text{Tr}\left[\int \frac{d^Dk}{(2\pi)^D}i\gamma^{a}\sigma_{bc}\Delta^{-1}(k)_{~d}^{\mu}iM\delta_{~\nu}^{d}\Delta^{-1}(k+p+q)_{~e}^{\nu}iM\delta_{~\rho}^{e}\Delta^{-1}(k+q)_{~a}^{\rho}\right]=0.
\eal
\ee
And finally
\be \label{W2S1}
\bal
\\
&\parbox{180mm}{
\includegraphics[]{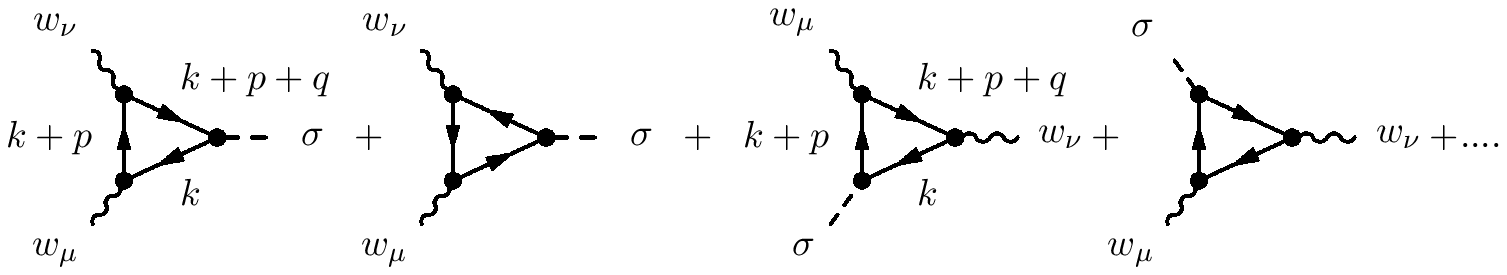}}
\\
=&-\frac{3~2!~2}{3!\cdot 2}\text{Tr}\left[\int \frac{d^Dk}{(2\pi)^D}i\gamma^{a}\sigma_{bc}\Delta^{-1}(k)_{~d}^{\mu}\gamma^ {d}\sigma_{ef}\Delta^{-1}(k+p+q)_{~g}^{\nu}iM\delta_{~\rho}^{g}\Delta^{-1}(k+p)_{~a}^{\rho}\right]\\
&-\frac{3~2!~2}{3!\cdot 2}\text{Tr}\left[\int \frac{d^Dk}{(2\pi)^D}i\gamma^{a}\sigma_{bc}\Delta^{-1}(k)_{~d}^{\mu}i\gamma^{d}\sigma_{ef}\Delta^{-1}(k+p+q)_{~g}^{\nu}iM\delta_{~\rho}^{g}\Delta^{-1}(k+q)_{~a}^{\rho}\right]\\
=&\left(\frac{2}{\epsilon}-\log{\left(\frac{M^2}{4 \pi\mu^2 }\right)}-\gamma -\frac{3}{4}\right)\left[-\frac{M^2}{4\pi^2}D^{\mu\nu}_{~~~ bcef}\right]-\frac{M^2}{16 \pi^2 }E^{\mu\nu}_{~~~ bcef}+\mathcal{O}(p^2).
\eal
\ee
With $D^{\mu\nu}_{~~~ bcef}$ and $E^{\mu\nu}_{~~~ bcef}$ being the same as in (\ref{W2})

\section{Summary of divergences}

In the previous section we obtained the results of the one-, two- and three-point 
functions for a general diagonal perturbation, sometimes
particularizing to a conformally flat metric to ease the notation. Let us now summarize the results.

The divergent part of diagrams (\ref{S0}-\ref{S03}) together with the $M^4$ piece of diagram (\ref{sigsig}) add up in the effective action to\footnote{Note that factors $1/n!$, where $n$ is the number of identical external legs, and a sign flip, are needed to reconstruct the term in the effective action from the diagrammatic calculation. }
\be \label{cosmo}
\frac{M^4e^{-\frac{\sum_i\sigma_i}{2}}}{8\pi^2}\left(\frac{2}{\epsilon}-\log{\left(\frac{M^2}{\mu^2 }\right)}\right).
\ee 
Note that this term's dimensionality matches (\ref{sqrg}). Furthermore, it can be proved that 
the divergent terms in (\ref{sigsig}) proportional to $M^2 p^2$ are precisely those corresponding to (\ref{Rmulti}) 
in momentum space, thus allowing us to recover the first orders of $\mathcal{L}_{R}$ for the general diagonal 
perturbation, which on shell correspond to $\sqrt{g}R$.

Diagram (\ref{W2}) has two divergent terms and neither seem to immediately correspond to any of the
possible counterterms discussed so far. The first one, that corresponding to $D^{\mu\nu}_{~~~ bcef}$, is 
of order $M^2\sqrt{g}R$. The other, 
corresponding to $F^{\mu\nu}_{~~~ bcef}$ is of higher order, i.e. $\mathcal{O}(\sqrt{g}R^2)$. Before addressing
these apparent new divergences let us particularize to the
case of a conformally flat perturbation above the vacuum. 
Taking $\sigma_{i}=\sigma$, (\ref{S0}-\ref{S03}) plus (\ref{sigsig}) add up in the effective action to 
\be \label{cosmo1}
\frac{M^4e^{-2\sigma}}{8\pi^2}\left(\frac{2}{\epsilon}-\log{\left(\frac{M^2e^{-\sigma}}{4 \pi\mu^2 }\right)}-\gamma +\frac{3}{2}\right)=\frac{M^4e^{-2\sigma +\frac{\sigma}{2}\epsilon}}{8\pi^2}\left(\frac{2}{\epsilon}-\log{\left(\frac{M^2}{4 \pi\mu^2 }\right)}-\gamma +\frac{3}{2}\right).
\ee 
An important remark is in order at this point. Note the peculiar form of $e^{-2\sigma +\frac{\sigma}{2}\epsilon}$: 
this factor corresponds to the determinant of a conformally flat metric in $D=4-\epsilon$ dimensions 
and it is a remanent of the fact that we used dimensional regularization to calculate the momentum integrals. 
Of course $\lim_{\epsilon\rightarrow 0}\sqrt{g_{D}}=\sqrt{g}$ (where $g_{D}$ is the determinant of 
the $D$-dimensional metric), but this is telling us that in order to regularize our integrals it is not 
enough to add a mass scale to match the dimensionality; an $\epsilon$ power of the determinant of 
the metric is also needed to ensure diffeomorphism invariance. That is $\mu^2\rightarrow \mu^2e^{-\sigma}$. 
Then (\ref{cosmo1}) would read
\be \label{cosmo2}
\frac{M^4e^{-2\sigma}}{8\pi^2}\left(\frac{2}{\epsilon}-\log{\left(\frac{M^2}{4 \pi\mu^2 }\right)}-\gamma +\frac{3}{2}\right).
\ee
Continuing with the divergences, diagrams (\ref{sigsig2}) and (\ref{sig3pots}) contain 
terms of order $M^4$ that are the subsequent 
orders of the expansion of (\ref{cosmo1}) in terms of $\sigma$. As for the terms of order $p^2$, one has to 
express them in position space. The result of diagram (\ref{sigsig}) for instance is
\be \label{pos}
-\frac{\sigma p^2\sigma M^2}{16\pi^2}\left(\frac{2}{\epsilon}-\log{\left(\frac{M^2}{4 \pi }\right)}-\gamma -\frac{2}{3}\right)
\ee
that in position space reads
\be
 \frac{\sigma\Box\sigma M^2}{16\pi^2}\left(\frac{2}{\epsilon}-\log{\left(\frac{M^2}{4 \pi }\right)}-\gamma -\frac{2}{3}\right).
\ee
The next diagrams we consider are (\ref{sigsig2}) plus (\ref{sig3pots})
\be \label{pos2}
\frac{M^2\sigma^2(p^2+(p+q)^2+q^2)\sigma}{32\pi^2}\left(\frac{2}{\epsilon}-\log{\left(\frac{M^2}{4 \pi }\right)}-\gamma -\frac{5}{3}\right),
\ee
or in position space
\be
 -\frac{3M^2\sigma^2\Box\sigma}{32\pi^2}\left(\frac{2}{\epsilon}-\log{\left(\frac{M^2}{4 \pi }\right)}-\gamma -\frac{5}{3}\right).
\ee
Now it is clear that the full calculation at order $M^2p^2$ resums to the following term in the effective action
\be \label{Ref}
\frac{M^2\Box\sigma e^{-\sigma}}{32\pi^2}\left(\frac{2}{\epsilon}-\log{\left(\frac{M^2}{4 \pi\mu^2 }\right)}-\gamma +\frac{1}{3}\right).
\ee
Note that this term has the same structure that $\sqrt{g}R$ for a conformally flat metric. This divergence
can be absorbed by redefining ${\cal L}_R$ and using the equations of motion.
This is already telling us that the theory is renormalizable only on shell\footnote{Please note that this is quite 
unrelated to the well known fact that pure gravity at one loop is finite on-shell. The latter result corresponds to 
performing a one-loop calculation with gravitons. Here instead we integrate the microscopic degrees of freedom 
that supposedly generate the gravitons after spontaneous symmetry breaking and generation of the metric degrees of freedom.}.
Namely when the spin connection  $w_{\mu}^{ab}$ corresponds to 
the Levi-Civita one. In our approach this identification is forced by the use of the equations of motion.

For a general diagonal perturbation one has to consider the momentum dependent $O(p^2)$ divergent pieces in
(\ref{sigsig}) and similar diagrams with more external scalar legs. As a check we can see that the momentum dependent
terms with two $\sigma_i$ fields faithfully reproduce the $O(\sigma^2)$ piece in the
curvature term (\ref{Rmulti}) thus confirming the general covariance of the effective action. 
Details are relegated to the Appendix.   

Let us now retake the issue of the apparent new divergences emerging from (\ref{W2}). To see if 
they really contribute to the final effective action we have to express them in terms 
of the $\sigma$ fields using the available equations of motion. Then in principle, they must 
either vanish or correspond to a valid counterterm. We argued in Section 2 that there is a 
third possible counterterm in 4D, the Gauss-Bonnet term, which is a total derivative and should not contribute to the dynamics. 
But it is reasonable that the higher order divergence in (\ref{sigsig}) constitutes a 
piece of that term that will ultimately cancel. Full details of this calculation can be found 
in the appendix showing how the lower order divergence vanishes and how the higher order term indeed 
corresponds to a piece of the Gauss-Bonnet term, and in particular vanishes for the conformally 
flat parametrization of the metric.

\section{Effective action and physical constants}

We are now ready to write the effective action we obtain on shell; that is once the spin connection is set to 
the value obtained after use of the equations of motion and the gap equation is used. 
We shall present details only for a vierbein corresponding to a
conformally flat metric but
as previously discussed we have a good check of its validity for the divergent parts 
of a general diagonal perturbation above the vacuum. 

We recall our conventions. We have used euclidean conventions so that the (emerging) metric has signature (+,+,+,+). The effective
action at long distances is defined by the functional integral 
\be
\int  [dg] \exp \left(- S[g]\right),
\ee
where 
\be
g_{\mu\nu}= \eta_{ab}e_\mu^a e_\nu^b = \frac{1}{M^2}\eta_{ab}B_\mu^a B_\nu^b,
\ee
according to our discussion in Section 2.

The effective action obtained after the diagrammatic calculation of the previous sections is
\be \label{efact} 
\bal
S_{eff}=&\int ~d^ 4x~\left(c'M^4e^{-2\sigma}-N\frac{M^4}{8\pi^2}e^{-2\sigma}\left(\log\left(\frac{M^2}{\mu^2}\right)-
\frac{3}{2}\right)\right. \\
&\left. +A'M^2\Box \sigma e^{-\sigma}-N\frac{M^2}{32\pi^2}\Box \sigma e^{-\sigma}\left(\log\left(\frac{M^2}{\mu^2}
\right)-\frac{28}{3}\right)\right)+...,
\eal
\ee
where $c'=c+\frac{N}{8\pi^2}\left(\frac{2}{\epsilon}+\log{4\pi}-\gamma\right)$ and 
$A'=A+\frac{N}{8\pi^2}\left(\frac{2}{\epsilon}+\log{4\pi}-\gamma\right)$ are renormalized coupling constants that have absorbed 
the divergences. The $\overline{MS}$ subtraction scheme is assumed. Note that the finite part of the term proportional to $M^2$ has received a contribution from the diagrams containing only $w_{\mu}^{ab}$ fields, see the appendix. Making use of the gap equation, 
we can write the previous expression as
\be \label{efact2} 
S_{eff}=\int ~d^4x~\left(N\frac{M^4}{16\pi^2}e^{-2\sigma}+A'\Box \sigma e^{-\sigma}M^2-N\frac{M^2}{32\pi^2}\Box \sigma e^{-\sigma}
\left(\log\left(\frac{M^2}{\mu^2}\right)-\frac{28}{3}\right)\right)+...,
\ee
The resulting effective theory thus describes a geometry with a cosmological term.
Sometimes it is stated in the literature\cite{tomboulis} that if gravity is an emergent phenomenon
and gravitons are Goldstone bosons all interactions should be of a derivative nature and the cosmological 
constant problem would
be in a sense solved. This is not so, as we see a cosmological terms is generated necessarily (both in 2D and 4D), at least
in the present approach.

The previous result is not exact of course. The effective action is in fact an infinite series
containing higher order derivatives, starting with terms of $O(\sqrt{g}R^2)$ and so on, which are represented by the
dots in the previous expression. In fact, as we have seen, a counterterm proportional to Gauss-Bonnet 
(of order ${\cal O}(\sqrt{g}R^2)$) is required; finite terms will appear too. The effective action
should also contain 
a non-local finite piece corresponding to the conformal anomaly (of dimension four in 4D\cite{anomaly}). The conformal 
anomaly was indeed reproduced in the previous work in 2D \cite{2D}. Note that any  dimension four term 
that is generated will be accompanied
by a factor of $N$. The dimension six terms will be of $O(N/M^2)$ and so forth. It would be natural to 
redefine the constant $A^\prime$ to
include this factor of $N$ in order to keep the counting of powers of $N$ homogeneous.

Appealing to covariance arguments we can now express (\ref{efact}) in terms of invariants
\be \label{efact2}
S_{eff}=\int ~d^ 4x~\left[\frac{N}{16\pi^2}M^4\sqrt{g}+
\left(A'-\frac{N}{48\pi^2}\left(\log\left(\frac{M^2}{\mu^2}\right)-\frac{28}{3}\right)\right)M^2\sqrt{g}R +...\right].
\ee

Next we recall that the classical Einstein action corresponding to the euclidean conventions is\cite{duff}
\be
S= - \frac{M_{P}^2}{32\pi}\int d^4 x \sqrt{g} (R-2\Lambda).
\ee
Now identifying
\be \label{cts}
\bal
\frac{N}{16\pi^2}M^4&=2\Lambda\frac{M_{P}^2}{32\pi}\\
M^2\left(A'-\frac{N}{48\pi^2}\left(\log\left(\frac{M^2}{\mu^2}\right)-\frac{28}{3}\right)\right)&=-\frac{M_{P}^2}{32\pi},
\eal
\ee
we indeed obtain
\be \label{efactEH}
S_{eff}=-\frac{M_{p}^2}{32\pi}\int ~d^ 4x~\sqrt{g}(R-2\Lambda)+ {\cal O}(p^4).
\ee
As we see from the previous discussion, the integration of the fermions (assumed to be the fundamental degrees of freedom
in the theory) yields a positive cosmological constant. As for the value of $M_P^2$, the Planck mass squared, the sign is 
not really automatically defined. More on this latter.

\subsection{Fine-tuning and running of the constants}

To ensure that the action is renormalization group invariant, thus observable, the following  
beta function for each free constant in the theory must be obeyed
\be \label{beta}
\bal
\mu \frac{dc'}{d\mu}&=-\frac{N}{4\pi^2}\\
\mu \frac{dA'}{d\mu}&=-\frac{N}{24\pi^2}\\
\eal
\ee
This running has nothing to do with the one generated by graviton exchange and it is 
thus unrelated to the presence or absence
of  asymptotic safety that some authors advocate for gravity. At scales $\mu \gg M$ the relevant degrees of 
freedom are not gravitons, but the $2N$ fermions appearing in the microscopic Lagrangian. On the other hand, 
at the moment that fermions become the 
relevant degrees of freedom, geometry loses its meaning. There is then no `shorter' distance 
than $M^{-1}$, or at the very least this regime 
cannot be probed. Note that to realize our physical assumption of having the fermions
as fundamental degrees of freedom we should have $c>0$ as discussed in Section 2.

These equations do not reflect the complete running of the dimensionless couplings associated to ${\cal L}_D$ and 
${\cal L}_R$, i.e. the constants associated to the cosmological and Einstein-Hilbert terms, but only the one obtained 
at leading order in $N$. That is, the `graviton' loops are not included here; they are suppressed by 
one power of $N$ if $N$ is large.
To see this last statement we recall that the usual power counting rules show that the exchange of the vierbein 
degrees of freedom would be accompanied
by a factor of $M_P^{-2}$, suppressed by $1/N$. Leaving these corrections aside, we note that the two free 
couplings of the theory 
have a running that is opposed in sign to the one found in 2D.

To understand the issue it is probably useful to appeal to the QCD analogy. 
At long distances strong interactions are well described
by the pion chiral Lagrangian, parametrized by $f_\pi$ or the $O(p^4)$ coefficients, 
generically named low energy
constants (LEC). The LEC are a complicated function of $\alpha_s$, the coupling constant of QCD. The microscopic 
theory proposed in this paper is the analogous of QCD, while the resulting effective
theory (\ref{efactEH}) is the counterpart of the chiral effective Lagrangian. Then $M_P$ and $\Lambda$ are the LEC of the present theory.
The running of $\alpha_s$  does not have an immediate translation on the LEC while in the present model, because of its simplicity, the 
consequences of the running in the microscopic particle reflects directly in $\Lambda$ and $M_P$. But in addition these
constants have an additional running (analogous to doing pion loops in the chiral Lagrangian). The counting of powers of $N$ 
disentangles both types of running.

At some scale, $q\sim M$ the effective theory stops making sense. At that moment the 
relevant degrees of freedom change and, as a result, the metric disappears. 
Exactly in the same way as for large momentum transfers we 
do not see pions but quarks.
Of course, if there is no metric there is no geometry and, in particular, the notion of distance disappears 
altogether at length scales below $M^{-1}$.
From this point of view, gravity is non-Wilsonian.

Let us now try to make contact with the value that the LEC take in gravity. Clearly, there is 
enough freedom in the theory (by adjusting
$A^\prime$ and $M$) to reproduce any values of $\Lambda$ and $M_P$. But we also want higher order terms 
to be small for the effective theory to
make sense in a reasonable range of momenta. We may even get rid of 
all of the high order (${\cal O}(p^4)$ and beyond) if we take $M\to \infty$ and at the same time we take $N\to 0$ in a prescribed form. 
Then, in the actual limit, which 
corresponds to a `quenched' approximation, we exactly reproduce
Einstein-Hilbert Lagrangian, with a cosmological constant, and nothing else. 
Of course in this limit, the presumed fundamental degrees of
freedom disappear completely and we have all the way up to $\mu=\infty$ Einstein's theory ---with all 
its ultraviolet problems; there are no fundamental degrees of freedom providing form factors to cut off 
the offending divergent integrals.

Of course the $N\to 0$ limit is just the opposite one to the one we have used. All our diagrammatic 
results are exact in the $N\to \infty$ limit and presumably get large corrections as $N$ approaches zero, 
but the general features of the model should survive. 

Note that $M$ is a fixed quantity in the model and if $M_P^2$ increases, $\Lambda$ decreases. Taking the actual 
observed or estimated values of these two parameters we get the value $N M^4\sim 10^{18}$ m$^{-4}$, which is a very low 
scale. One may think 
that this may already represent unacceptably large corrections from higher order operators. However, this is not
necessarily so because the bounds 
on $R^2$ terms are very weak. For instance, the bound $ k < 10^{74}$ has been quoted for 
a generic coefficient\cite{stelle} $k$ of the ${\cal O}(p^4)$ terms. Thus, a relatively low scale 
for $M$ cannot be really excluded observationally 
by studying gravitational effects alone and one 
should be aware of this. However, our own perception tells us that $M$ should be much larger than the value quoted
above as the notion of 
metric certainly makes sense at much shorter distances. We 
can increase the value of $M$ as much as we want by decreasing the value of  $N$, 
as previously indicated. We shall not elaborate further on this as it seems too premature an speculation.

Finally we note that the sign of Newton's constant is not determined a priori in 
this theory due to the subtraction required from the
counterterm in $\cal{L}_R$. This ties nicely with some of the early discussions on induced gravity\cite{adler}.

\section{Summary}

Let us summarize our findings and comment on possible implications of our work. We have proposed a model 
where 4D gravity emerges from a theory
without any predefined metric. The minimal input is provided by assuming a differential 
manifold structure endowed with
an affine connection. Nothing more. The Lagrangian can be defined without having to appeal
to a particular metric or vierbein.

Gravity and
distance are induced rather than fundamental concepts in this proposal. At sufficiently short scales, when the
effective action does not make sense anymore, the physical degrees of freedom are fermionic.
At such short scales there is not even the notion of a smaller scale.  

The relative technical simplicity of this proposal constitutes its main virtue when 
compared with previous proposals \cite{ru,wett}, where
even semiquantitative discussions appear impossible. Here one is able to derive in full detail
the effective action.

A very important aspect of the model is the improvement of the ultraviolet behavior. 
After integration of the fundamental degrees of freedom all the divergences
can be absorbed in the redefinition of the cosmological constant and the Planck mass (as seen from
the effective theory point of view, even though the respective counterterms do not have this meaning in 
the underlying theory. With the running,
dictated by the corresponding beta functions, both quantities are renormalization group invariant. In addition
the Gauss-Bonnet invariant is renormalized too. This happens in spite of the bad ultraviolet behavior of 
the propagator and the ultimate reason, we think, is that these are the {\em only} counterterms that
can be written without having to assume an underlying metric that does not exist before spontaneous
symmetry breaking takes place.

At long distances the fluctuations around the broken vacuum are the relevant degrees of freedom and are described
by an effective theory whose lowest dimensional operators are just those of ordinary 4D gravity. They
of course exhibit the usual divergences of quantum gravity but this now poses in principle no problem as we know
that at very high energies this is not the right theory. For $q\sim M$ one starts seeing the fundamental degrees of freedom.
Gravitons are the Goldstone bosons of a broken global symmetry. We have seen how the barrier of the 
Weinberg-Witten no-go theorem could be overcome.

In a sense the fundamental fermions resolve the point-like 3-graviton, 4-graviton, etc.  interactions into 
extended form factors and this is the reason for the mitigation of the terrible ultraviolet behavior of 
quantum gravity. However this is only part of the story, because this could be equally achieved by using 
Dirac fermions coupled to gravity (or any other field for that matter). This would in fact be just a reproduction
of the old program of induced gravity\cite{adler} and therefore not that interesting. The really novel point in
this proposal is that the microscopic fermion action does not contain any metric tensor at all. Then not only is 
the metric and its fluctuations --the gravitons-- spontaneously generated, but the possible counterterms are
severely limited in number.

We stop short of making any strong claims about the renormalizability of the model. We can just say that, from 
our calculations and our experience with the model, renormalizability is a plausible hypothesis (our results
actually amount to an heuristic proof in the large $N$ limit). Likewise we do not insist in that the one
presented is the {\em sole} possibility to carry out the present program, although it looks fairly unique.
Clearly a number of issues need further study before the present proposal can be taken seriously but we 
think that the results presented here are of sufficient interest to make them known to researchers
in the field.

A number of extensions and applications come to our mind. Perhaps the most intriguing one from a physical point of view
would be to investigate in this framework singular solutions in GR such as black holes. A more in-depth 
study of the renormalizability issue is certainly required too as there are issues related to the renormalization group to be addressed
in the present setting.

\section*{Acknowledgements}
We acknowledge the financial support of projects
FPA2010-20807 and SGR2009SGR502. This research is supported by the Consolider CPAN
project. The work of J.A. is partially supported by
VRAID/DID/46/2010 and Fondecyt 1110378. D.E. and D.P. would like to thank the Faculty of Physics of PUC and the CERN PH-TH Unit, 
where part of this research was done, for hospitality.

\newpage

\appendix
\numberwithin{equation}{section}
\section{Appendix}

In this appendix we include the explicit calculation of the different terms appearing in (\ref{W2}) showing 
how they correspond on shell to different terms in the action. We also include, for completeness, how the 
result of (\ref{sigsig}) in the diagonal parametrization of the metric used in the text yields precisely (\ref{Rmulti}).

\subsection{$D^{\mu\nu}_{~~~ bcef}$, $E^{\mu\nu}_{~~~ bcef}$ and $F^{\mu\nu}_{~~~ bcef}$}
We saw in subsection (4.2) that diagram (\ref{W2}) contains three different terms, two of them being divergent, 
let us show how they either cancel or can be accommodated in the available counterterms. Let us write them 
down together with the $w_{\mu}^{ab}$ fields.
\be \label{D}
\bal
D^{\mu\nu}_{~~~ bcef}w_{\mu}^{bc}w_{\nu}^{ef}=&4w_{\mu}^{\nu b}w_{\nu}^{\mu b}-2w_{\mu}^{be}w_{\mu}^{be}=2(w_{1}^{12})^2-2(w_{1}^{21})^2+2(w_{1}^{13})^2-2(w_{1}^{31})^2\\
&+2(w_{1}^{14})^2-2(w_{1}^{41})^2+...+2(w_{4}^{34})^2-2(w_{4}^{43})^2=0.
\eal
\ee
So this divergence cancels regardless of the parametrization we choose. 

Let us write the second one now considering a conformally flat parametrization of the metric
\be \label{flat}
g_{\mu\nu}= e^{-\sigma(x)}\delta_{\mu\nu}.
\ee
Then we have
\be \label{E}
\bal
E^{\mu\nu}_{~~~ bcef}w_{\mu}^{bc}w_{\nu}^{ef}=4w_{\mu}^{\mu b}w_{\nu}^{b \nu}
\eal
\ee
Making use of equation (\ref{eom2}), that is 
\be \label{eoma}
 w_{\mu}^{~ab}=\frac{1}{2}(\partial^{a}\sigma\delta_{\mu}^{b}-\partial^{b}\sigma\delta_{\mu}^{a}),
\ee
we obtain
\be \label{E2}
\bal
E^{\mu\nu}_{~~~ bcef}w_{\mu}^{bc}w_{\nu}^{ef}=-9[(\partial_{1}\sigma)^2+(\partial_{2}\sigma)^2+(\partial_{3}\sigma)^2+(\partial_{4}\sigma)^2]=-9\partial_{\mu}\sigma\partial_{\mu}\sigma.
\eal
\ee
Recall this term appeared both in (\ref{W2}) and in (\ref{W2S1}). Summing their contributions in the way we 
did to reconstruct the effective action for the $\sigma$ field diagrams we obtain
\be \label{Ra}
\frac{9M^2}{32\pi^2}\partial_{\mu}\sigma\partial_{\mu}\sigma(1-\sigma+...)\rightarrow \frac{9M^2}{32\pi^2}\partial_{\mu}\sigma\partial_{\mu}\sigma e^{-\sigma}\rightarrow \frac{9M^2}{32\pi^2}e^{-\sigma}\Box \sigma.
\ee
This is just a finite contribution to $\sqrt{g}R$ and it is reflected in (\ref{efact})

The Gauss-Bonnet term corresponding to such a metric perturbation reads
\be \label{gb}
\bal
\mathcal{L}_{GB}&=\sqrt{g}(R^2+4R_{\mu\nu}R^{\mu\nu}-R_{\mu\nu\rho\sigma}R^{\mu\nu\rho\sigma})\\
&=-4\partial_{3}\partial_{4}\sigma\partial_{3}\partial_{4}\sigma+4\partial_{4}^{2}\sigma\partial_{3}^{2}\sigma-4\partial_{2}\partial_{4}\sigma\partial_{2}\partial_{4}\sigma+4\partial_{4}^{2}\sigma\partial_{2}^{2}\sigma-4 \partial_{2}\partial_{3}\sigma\partial_{2}\partial_{3}\sigma+4\partial_{3}^{2}\sigma\partial_{2}^{2}\sigma\\
&-4\partial_{1}\partial_{4}\sigma\partial_{1}\partial_{4}\sigma+4\partial_{4}^{2}\sigma\partial_{1}^{2}\sigma-4\partial_{1}\partial_{2}\sigma\partial_{1}\partial_{2}\sigma+4\partial_{2}^{2}\sigma\partial_{1}^{2}\sigma-4\partial_{1}\partial_{3}\sigma\partial_{1}\partial_{3}\sigma+4\partial_{3}^{2}\sigma\partial_{1}^{2}\sigma\\
&-3\partial_{4}\sigma\partial_{4}\sigma\partial_{4}^{2}\sigma-\partial_{4}^{2}\sigma\partial_{3}\sigma\partial_{3}\sigma-4\partial_{4}\sigma\partial_{3}\sigma\partial_{3}\partial_{4}\sigma-\partial_{4}\sigma\partial_{4}\sigma\partial_{3}^{2}\sigma-3\partial_{3}\sigma\partial_{3}\sigma\partial_{3}^{2}\sigma-\partial_{4}^{2}\sigma\partial_{2}\sigma\partial_{2}\sigma\\
&-\partial_{3}^{2}\sigma\partial_{2}\sigma\partial_{2}\sigma-4\partial_{4}\sigma\partial_{2}\sigma\partial_{2}\partial_{4}\sigma-4\partial_{3}\sigma\partial_{2}\sigma \partial_{2}\partial_{3}\sigma-\partial_{4}\sigma\partial_{4}\sigma\partial_{2}^{2}\sigma-\partial_{3}\sigma\partial_{3}\sigma\partial_{2}^{2}\sigma-3\partial_{2}\sigma\partial_{2}\sigma\partial_{2}^{2}\sigma\\
&-\partial_{4}^{2}\sigma\partial_{1}\sigma\partial_{1}\sigma-\partial_{3}^{2}\sigma\partial_{1}\sigma\partial_{1}\sigma-\partial_{2}^{2}\sigma\partial_{1}\sigma\partial_{1}\sigma-4\partial_{4}\sigma\partial_{1}\sigma\partial_{1}\partial_{4}\sigma-4\partial_{3}\sigma\partial_{1}\sigma\partial_{1}\partial_{3}\sigma-4\partial_{2}\sigma\partial_{1}\sigma\partial_{1}\partial_{2}\sigma\\
&-\partial_{4}\sigma\partial_{4}\sigma\partial_{1}^{2}\sigma-\partial_{3}\sigma\partial_{3}\sigma\partial_{1}^{2}\sigma-\partial_{2}\sigma\partial_{2}\sigma\partial_{1}^{2}\sigma
-3\partial_{1}\sigma\partial_{1}\sigma\partial_{1}^{2}\sigma.
\eal
\ee
The last term we have to explore is the piece $\frac{1}{\epsilon}F^{\mu\nu}_{~~~ bcef}$. Let us explicitly 
write this term together with the $w_{\mu}^{bc}$ fields
\be \label{a4}
\bal
\frac{1}{\epsilon}F^{\mu\nu}_{~~~ bcef}w_{\mu}^{bc}w_{\nu}^{ef}=&\frac{1}{\epsilon}\left[-\frac{\delta^{\mu\nu} \delta_{be}p_{c}p_{f}}{12 \pi ^2}+\frac{\delta^{\mu\nu} \delta_{bf}p_{e} p_{c}}{12 \pi ^2}+\frac{\delta^{\mu\nu} \delta_{ce}p_{b} p_{f}}{12 \pi ^2}-\frac{\delta^{\mu\nu}\delta_{cf}p_{b} p_{e}}{12 \pi ^2}-\frac{\delta^{\mu}_{ b} \delta^{\nu}_{ e}\delta_{cf} p^2}{12 \pi ^2}\right. \\
 &\left. +\frac{\delta^{\mu}_{ b} \delta^{\nu}_{ e}p_{c}p_{f}}{12 \pi ^2}+\frac{\delta^{\mu}_{ b} \delta^{\nu}_{ f} \delta_{ce} p^2}{12 \pi ^2}-\frac{\delta^{\mu}_{ b} \delta^{\nu}_{ f} p_{e} p_{c}}{12 \pi ^2}-\frac{\delta^{\mu}_{ e} \delta_{bf}p^{\nu} p_{c}}{12 \pi ^2}+\frac{\delta^{\mu}_{ e} \delta_{b }^{\nu}p_{c}p_{f}}{12 \pi ^2}\right.\\
 &\left. -\frac{\delta^{\mu}_{ e} \delta_{c }^{\nu}p_{b} p_{f}}{12 \pi ^2}+\frac{\delta^{\mu}_{ e}\delta_{cf}p^{\nu}p_{b}}{12 \pi ^2}+\frac{\delta^{\mu}_{ c} \delta^{\nu}_{ e} \delta_{bf} p^2}{12 \pi ^2}-\frac{\delta^{\mu}_{ c} \delta^{\nu}_{ e}p_{b} p_{f}}{12 \pi ^2}-\frac{\delta^{\mu}_{ c} \delta^{\nu}_{ f} \delta_{be} p^2}{12 \pi ^2}\right.\\
 &\left. +\frac{\delta^{\mu}_{ c} \delta^{\nu}_{ f}p_{b} p_{e}}{12 \pi ^2}+\frac{\delta^{\mu}_{ f} \delta_{be}p^{\nu} p_{c}}{12 \pi ^2}-\frac{\delta^{\mu}_{ f} \delta_{b}^{ \nu} p_{e} p_{c}}{12 \pi ^2}+\frac{\delta^{\mu}_{ f} \delta_{c }^{\nu}p_{b} p_{e}}{12 \pi ^2}-\frac{\delta^{\mu}_{ f} \delta_{ce}p^{\nu}p_{b}}{12 \pi ^2}\right. \\
&\left. +\frac{\delta_{be} \delta_{c}^{ \nu} p^{\mu} p_{f}}{12 \pi ^2}-\frac{\delta_{be}\delta_{cf} p^{\mu}p^{\nu}}{12 \pi ^2}-\frac{\delta_{bf} \delta_{c}^{ \nu} p^{\mu} p_{e}}{12 \pi ^2}+\frac{\delta_{bf} \delta_{ce} p^{\mu}p^{\nu}}{12 \pi ^2}-\frac{\delta_{b}^{ \nu} \delta_{ce} p^{\mu} p_{f}}{12 \pi ^2}\right. \\
 &\left. +\frac{\delta_{b}^{ \nu}\delta_{cf} p^{\mu} p_{e}}{12 \pi ^2}\right]w_{\mu}^{bc}w_{\nu}^{ef}\\
 =&\frac{1}{ 3\pi^ 2}\frac{1}{\epsilon}\left[\partial_{c}w_{\mu}^{bc}\partial_{e}w_{\mu}^{be}+w_{\mu}^{\mu c}\Box w_{\nu}^{\nu c}-\partial_{c}w_{\mu}^{\mu c}\partial_{e}w_{\nu}^{\nu e}-\partial_{c}w_{\mu}^{bc}\partial_{e}w_{b}^{\mu e}\right.\\
 &\left. +\partial_{c}w_{\mu}^{bc}\partial^{\nu}w_{\nu}^{\mu b}+\partial^{\mu}w_{\mu}^{b\nu}\partial_{e}w_{\nu}^{e b}+\frac{1}{2}\partial^{\mu}w_{\mu}^{bc}\partial^{\nu}w_{\nu}^{bc}\right]
\eal
\ee

\be \label{a5}
\bal
\frac{1}{\epsilon}F^{\mu\nu}_{~~~ bcef}w_{\mu}^{bc}w_{\nu}^{ef}
 =\frac{6}{\pi^ 2}\frac{1}{\epsilon}&\left[ - \partial_{4}^2\sigma \partial_{1}^2\sigma+ \partial_{1}\partial_{4}\sigma\partial_{1}\partial_{4}\sigma
       -  \partial_{3}^2\sigma \partial_{4}^2\sigma + \partial_{3}\partial_{4}\sigma\partial_{3}\partial_{4}\sigma   \right. \\ 
       &\left. -  \partial_{2}^2\sigma \partial_{4}^2\sigma+ \partial_{2}\partial_{4}\sigma\partial_{2}\partial_{4}\sigma-  \partial_{2}^2\sigma \partial_{3}^2\sigma+ \partial_{2}\partial_{3}\sigma\partial_{2}\partial_{3}\sigma \right. \\
       &\left. -  \partial_{3}^2\sigma \partial_{1}^2\sigma+\partial_{1}\partial_{3}\sigma\partial_{1}\partial_{3}\sigma -  \partial_{2}^2\sigma \partial_{1}^2\sigma+ \partial_{1}\partial_{2}\sigma\partial_{1}\partial_{2}\sigma\right].
\eal
\ee
Now it is easy to see that (\ref{a5}) corresponds to the second and third lines in (\ref{gb}). This divergent 
contribution is part of the Gauss-Bonnet term, which, although being a total derivative, is a valid counterterm. 
The rest of (\ref{gb}) contains three sigma fields and is not present in (\ref{W2}) as it should be the case. 
These remaining terms would be generated in the triangular diagram with three external $w_{\mu}^{ab}$ fields and would come 
with a divergent coefficient (we have not computed such diagram). 
Note also that both (\ref{a5}) and the first two lines of (\ref{gb}) can be integrated by parts to make them vanish. 
This happens because in the conformally flat metric perturbations terms from the diagrams with two and three external $w$ 
fields can not be related to each other integrating by parts. Therefore, they must vanish independently as the Gauss-Bonnet 
term is a total derivative after all. 

This is a particularity of the conformally flat parametrization of the metric perturbation and would not hold for a 
generally diagonal parametrization. In that case terms generated in the two point function can be transformed 
into the terms appearing in the three point function by integration by parts and one would require a full 
calculation to show there is a match with an independent calculation of the Gauss-Bonnet term.

\subsection{$\sqrt{g}R$ for the general diagonal parametrization of the metric}
Let us consider a generally diagonal metric with four degrees of freedom
\be \label{a1}
g_{\mu\nu}=\left(\begin{matrix}
e^{-\sigma_{1}(x)} & 0 & 0 & 0\\
0 & e^{-\sigma_{2}(x)} & 0 & 0\\
0 & 0 & e^{-\sigma_{3}(x)} & 0\\
0 & 0 & 0 & e^{-\sigma_{4}(x)}\\
\end{matrix}\right).
\ee
We saw that the corresponding expression for the curvature is Eq. (\ref{Rmulti})
\be \label{a2}
\bal
\mathcal{L}_{R}|_{\text{(on shell)}}=&M^2\sqrt{g}R=M^2
\left[
 \partial_{3}^{2}\sigma_{4}+\partial_{2}^{2}\sigma_{4}+\partial_{1}^{2}\sigma_{4}
 +\partial_{4}^{2}\sigma_{3}+\partial_{2}^{2}\sigma_{3}+\partial_{1}^{2}\sigma_{3}
 \right.\\
&\left.+\partial_{4}^{2}\sigma_{2}+\partial_{3}^{2}\sigma_{2}+\partial_{1}^{2}\sigma_{2}+\partial_{4}^{2}\sigma_{1}+\partial_{3}^{2}\sigma_{1}+\partial_{2}^{2}\sigma_{1}
\right.\\
&\left.
 -\frac{1}{2}\left(\partial_{3}\sigma_{1}\partial_{3}\sigma_{2}+\partial_{4}\sigma_{1}\partial_{4}\sigma_{2}
+\partial_{2}\sigma_{1}\partial_{2}\sigma_{3}+\partial_{4}\sigma_{1}\partial_{4}\sigma_{3}+\partial_{2}\sigma_{1}\partial_{2}\sigma_{4}
+\partial_{3}\sigma_{1}\partial_{3}\sigma_{4}\right.\right.\\
&\left.\left. +\partial_{1}\sigma_{2}\partial_{1}\sigma_{3}+\partial_{4}\sigma_{2}\partial_{4}\sigma_{3}+\partial_{1}\sigma_{2}\partial_{1}\sigma_{4}+\partial_{3}\sigma_{2}\partial_{3}\sigma_{4}
+\partial_{1}\sigma_{3}\partial_{1}\sigma_{4}+\partial_{2}\sigma_{3}\partial_{2}\sigma_{4}\right)\right.\\
&\left. +\mathcal{O}(\sigma^3)\right]
.
\eal
\ee
We consider now the divergent part proportional to $M^2$ of the result of (\ref{sigsig})
\be \label{a3}
\bal
&\frac{2}{\epsilon}\sum_{j=1}^{4}\sum_{\bal l=1\\ l\neq j\eal}^{4}\left[-\frac{\sigma_{j}\sigma_{l}p^2M^2}{48\pi^2}
+\frac{\sigma_{j}\sigma_{l}(p_{j}^2+p_{l}^{2})M^2}{48\pi^2}\right]\\
=&\frac{2}{\epsilon}\frac{M^2}{48\pi^2}\left[-p^2\left(\sigma_{1}\sigma_{2}+\sigma_{1}\sigma_{3}+\sigma_{1}\sigma_{4}+\sigma_{2}\sigma_{3}+\sigma_{2}\sigma_{4}
+\sigma_{3}\sigma_{4}\right)\right.\\
&\left.+\sigma_{1}\sigma_{2}(p_{1}^2+p_{2}^2)+\sigma_{1}\sigma_{3}(p_{1}^2+p_{3}^2)+\sigma_{1}\sigma_{4}(p_{1}^2+p_{4}^2)+\sigma_{2}\sigma_{3}(p_{2}^2+p_{3}^2)\right.\\
&\left. +\sigma_{2}\sigma_{4}(p_{2}^2+p_{4}^2)+\sigma_{3}\sigma_{4}(p_{3}^2+p_{4}^2)\right]\\
=&\frac{2}{\epsilon}\frac{M^2}{48\pi^2}\left[-\sigma_{1}\sigma_{2}(p_{3}^2+p_{4}^2)-\sigma_{1}\sigma_{3}(p_{2}^2+p_{4}^2)-\sigma_{1}\sigma_{4}(p_{2}^2+p_{3}^2)-\sigma_{2}\sigma_{3}(p_{1}^2+p_{4}^2)\right.\\
&\left. -\sigma_{2}\sigma_{4}(p_{1}^2+p_{3}^2)-\sigma_{3}\sigma_{4}(p_{1}^2+p_{2}^2)\right].
\eal
\ee
This last expression, when expressed in position space, corresponds exactly to (\ref{a2}) except for a numerical factor and minus the second derivatives of the fields which are total derivatives and do not appear in the perturbative calculation.

\end{document}